\documentclass[12pt]{article}

\usepackage{sbc-template}
\usepackage{graphicx,url}
\usepackage[utf8]{inputenc}
\usepackage[english]{babel}
\usepackage{multirow}
\usepackage{graphicx}
\usepackage{float}
\usepackage{multirow}
\usepackage[nolist]{acronym}
\usepackage{amsmath}

\usepackage{tikz}
\newcommand{\circlednumber}[1]{%
    \tikz[baseline=(char.base)]{
        \node[shape=circle, draw=black, fill=black, text=white, inner sep=0.7pt, font=\sffamily\small] (char) {#1};
    }
}
     
\title{Resource Allocation Influence on Application Performance in Sliced Testbeds}

\author{Rodrigo Moreira\inst{1}, Larissa F. {Rodrigues Moreira}\inst{1,2}, Tereza C. Carvalho\inst{3}, \\ Flávio de Oliveira Silva\inst{2,4}}

\address{Federal University of Viçosa
  (UFV)\\
  38.810-000 -- Rio Paranaíba -- MG -- Brazil
\nextinstitute
  Faculty of Computing -- Federal University of Uberlândia (UFU)\\
  38.408-100 -- Uberlândia -- MG -- Brazil
\nextinstitute
  University of São Paulo (USP)\\
  05.508-010 -- São Paulo -- SP -- Brazil
\nextinstitute
  Department of Informatics -- School of Engineering\\
  University of Minho -- Braga, Portugal
  \email{\{rodrigo, larissa.f.rodrigues\}@ufv.br, flavio@di.uminho.pt,}  \email{terezacarvalho@usp.br, \{larissarodrigues, flavio\}@ufu.br}
}

\begin{document} 
\acrodef{3GPP}{3rd Generation Partnership Project}
\acrodef{AI}{Artificial Intelligence}
\acrodef{B5G}{Beyond Fifth Generation}

\acrodef{CPU}{Central Processing Unit}
\acrodef{DoS}{Denial of Service}
\acrodef{DDoS}{Distributed Denial of Service}
\acrodef{DNN}{Deep Neural Network}
\acrodef{DRL}{Deep Reinforcement Learning}
\acrodef{DT}{Decision Tree}
\acrodef{ETSI}{European Telecommunications Standards Institute}
\acrodef{FIBRE}{Future Internet Brazilian Environment for Experimentation}
\acrodef{FIBRE-NG}{Future Internet Brazilian Environment for Experimentation New Generation}
\acrodef{GNN}{Graph Neural Networks}
\acrodef{HTM}{Hierarchical Temporal Memory}

\acrodef{IAM}{Identity And Access Management}
\acrodef{IID}{Informally, Identically Distributed}
\acrodef{IoE}{Internet of Everything}
\acrodef{IoT}{Internet of Things}
\acrodef{KNN}{K-Nearest Neighbors}
\acrodef{LSTM}{Long Short-Term Memory}
\acrodef{M2M}{Machine to Machine}
\acrodef{MAE}{Mean Absolute Error}
\acrodef{ML}{Machine Learning}
\acrodef{MOS}{Mean Opinion Score}
\acrodef{MAPE}{Mean Absolute Percentage Error}
\acrodef{MSE}{Mean Squared Error}
\acrodef{mMTC}{Massive Machine Type Communications}
\acrodef{MFA}{Multi-factor Authentication}
\acrodef{MQTT}{Message Queuing Telemetry Transport}

\acrodef{OSM}{Open Source MANO}
\acrodef{QoE}{Quality of experience}
\acrodef{QoS}{Quality of Service}
\acrodef{RAM}{Random-Access Memory}
\acrodef{RF}{Random Forest}
\acrodef{RL}{Reinforcement Learning}
\acrodef{RMSE}{Root Mean Square Error}
\acrodef{RNN}{Recurrent Neural Network}
\acrodef{RAM}{Random-Access Memory}

\acrodef{SDN}{Software-Defined Networking}
\acrodef{SFI2}{Slicing Future Internet Infrastructures}
\acrodef{SLA}{Service-Level Agreement}
\acrodef{SON}{Self-Organizing Network}

\acrodef{VoD}{Video on Demand}
\acrodef{VR}{Virtual Reality}
\acrodef{V2X}{Vehicle-to-Everything}


\maketitle

\begin{abstract}

Modern network architectures have shaped market segments, governments, and communities with intelligent and pervasive applications. Ongoing digital transformation through technologies such as softwarization, network slicing, and AI drives this evolution, along with research into Beyond 5G (B5G) and 6G architectures. Network slices require seamless management, observability, and intelligent-native resource allocation, considering user satisfaction, cost efficiency, security, and energy. Slicing orchestration architectures have been extensively studied to accommodate these requirements, particularly in resource allocation for network slices. This study explored the observability of resource allocation regarding network slice performance in two nationwide testbeds. We examined their allocation effects on slicing connectivity latency using a partial factorial experimental method with \ac{CPU} and memory combinations. The results reveal different resource impacts across the testbeds, indicating a non-uniform influence on the CPU and memory within the same network slice.
  
\end{abstract}
     
\section{Introduction}\label{sec:introduction}

Network slicing enables logical, service-tailored, and independent networks to coexist in a shared physical network~\cite{Feng2020, Moreira2021, Donatti2023}. Network slicing allows application verticals with different \ac{SLA} to be orchestrated under heterogeneous underlying infrastructures. The optimal allocation of resources to network slices is fundamental for cost reduction, energy harvesting, and compliance with \ac{SLA}~\cite{Motalleb2023}.

Many efforts, such as combinatorial optimization and computational intelligence, have been aimed at effectively managing resource allocation for network slices~\cite{Debbabi2020}. Although approaches to resource allocation are still under development, understanding the behavior and observability of this allocation on network slicing performance still poses challenges~\cite{Saibharath2023}.

In this paper, we propose and evaluate the influence of \ac{CPU} and \ac{RAM} resource allocation on the performance of a network slicing application deployed on \ac{FIBRE-NG} and Fabric testbeds. Using the partial factorial performance evaluation method, we built resource allocation templates. We combined them for allocation to the network slice and measured the influence of the combination on the latency response variable for Write (W) and Read (R) operations.

The remainder of this paper is organized as follows. In Section~\ref{sec:related_work}, we contextualize the related work on testbed experimentation. The proposed experimental method is presented in detail in Section~\ref{sec:evaluation_proposal}, followed by a description of the experimental setup and results in Section~\ref{sec:results_and_discussions}. Section~\ref{sec:concluding_remarks} discusses concluding remarks and future research directions.

\section{Related Work}\label{sec:related_work}

In this section, we present related works concerning the deployment of network slices in experimental testbeds.

\cite{Dong2023} presents LinkLab 2.0, a multi-tenant IoT testbed with edge-cloud integration. The authors aim to address the challenges of programming and experimenting with heterogeneous IoT, edge, and cloud devices in a unified way. They design and implement a three-tiered architecture for managing the devices, a one-site programming framework for supporting serverless functions and computation offloading, and an anomaly detection system for ensuring reliability. They deploy LinkLab 2.0 with over 420 real devices of 14 types and support various research and educational experiments.

\cite{Morel2023} introduce a method for managing network services in Visual Cloud Computing (VCC) applications that use video streaming across edge-to-cloud systems. It utilizes P4-enabled programmable data planes and In-band Network Telemetry (INT) to boost video delivery quality and performance. Tested on the FABRIC network, it uses a customized P4 program merging Multi-Hop Route Inspection (MRI) and port forwarding to monitor and control congested network traffic. Results show improvements in packet loss, throughput, and delay compared to standard switches. 

\cite{Arora2024} presents a Cloud-native Lightweight Slice Orchestration (CLiSO), a framework for managing network slices using Kubernetes and a CISM agent. It emphasizes Domain Specific Handlers (DSHs) for automated network slicing management. The framework is evaluated by orchestrating OpenAirInterface functions on different cloud platforms, showcasing efficient orchestration, low resource use, resilience, and quick deployment.

\section{Evaluation Proposal}\label{sec:evaluation_proposal}

In this paper we designed and evaluated the performance of a network slicing application on real nationwide testbeds using partial factorial methodology. For this evaluation, we used the Cassandra application, a scalable, fault-tolerant, distributed key-value database system, for managing extensive data across multiple locations~\cite{padalia2015apache}. We deployed Cassandra in two experimental testbeds to measure the influence of allocating CPU and \ac{RAM} resources to the microservices of the network slice on the latency in read and write operations.

\subsection{Experimental Setup}\label{subsec:experimental_setup}

\begin{figure}[htp]
  \centering
  \includegraphics[width=0.9\columnwidth]{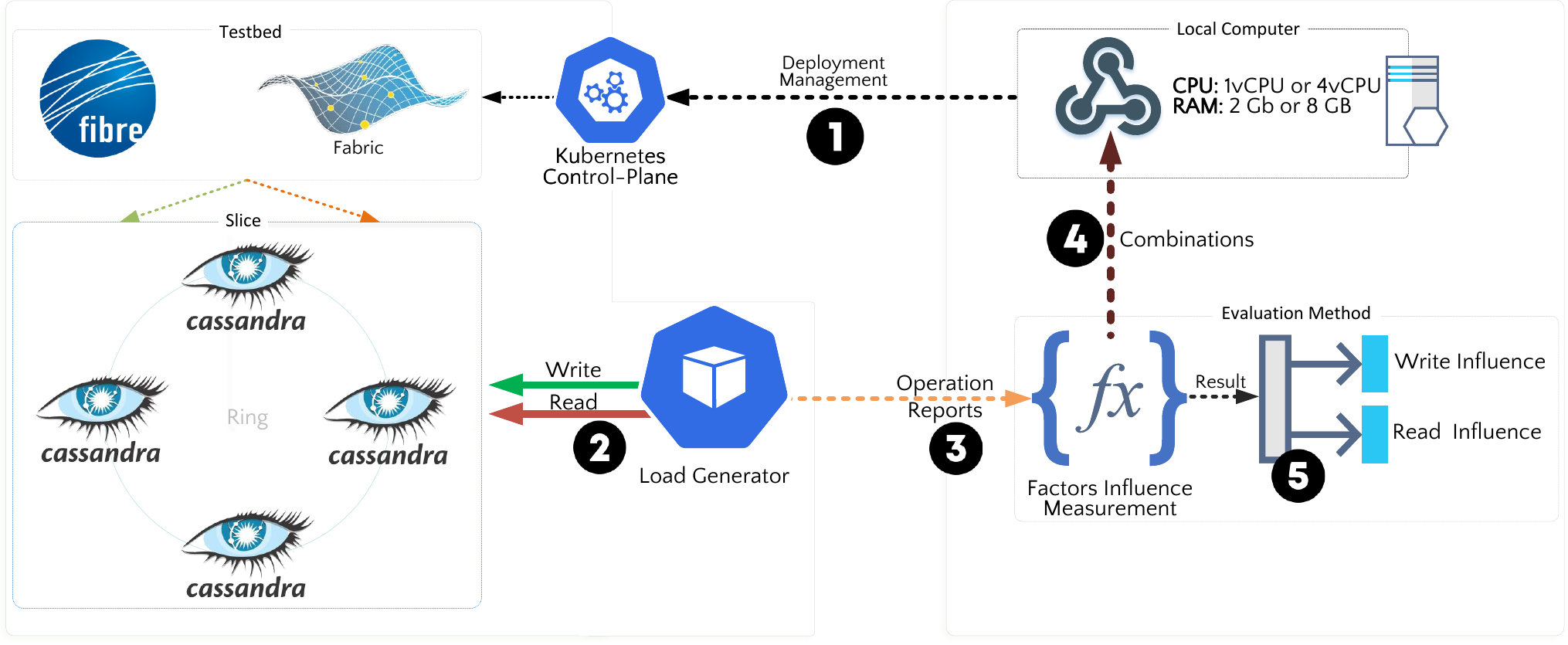}
  \caption{Evaluation Method.}
  \label{fig:method}
\end{figure}

We illustrate the evaluation method and the experimental flow and testbed in Figure~\ref{fig:method}. According to step one \circlednumber{1}, we deployed a Cassandra application on two different testbeds: \ac{FIBRE} New Generation and Fabric~\cite{salmito2014fibre, baldin2019fabric}. Our Cassandra application is based on the microservices model, in which different service parts (containers) run on different computing nodes of the testbed. The basic configuration of Cassandra in our experimental evaluation was three (3) replicas, each with 1024 data tokens.

In step two \circlednumber{2}, there is a workload generator container that triggers operations towards the Cassandra ring. Internally, the workload generator container is equipped with the \texttt{cassandra-stress} application, where we configure the operation parameters (W or R) such as time, data volume (10,000 entries), distribution (replication factor 2), and consistency level (\textit{quorum}). The workload generated on the Cassandra ring generates statistical outputs (operations per second, lines per second, latency and others). In step three \circlednumber{3} we use these statistics in our partial factorial influence method on the response variable latency.

Variations in resource allocation were combined using a local script and reorganized in step four \circlednumber{4}. In step five \circlednumber{5}, we determine the influence of the CPU and \ac{RAM} factors and their levels on the latency of W and R in different testbeds. In our experimental design, we built a slice on the \ac{FIBRE-NG} and Fabric testbeds considering the computational nodes presented in Table~\ref{tab:deployment_over_testbeds}.

\begin{table}[htp]
\centering
\scriptsize
\caption{Allocation of physical nodes to network slice service.}
\label{tab:deployment_over_testbeds}
\begin{tabular}{cccc}
\hline
\textbf{Testbed}                                       & \textbf{Pod Name} & \textbf{Management IP} & \textbf{Node} \\ \hline
\multicolumn{1}{l}{\multirow{4}{*}{\textbf{FIBRE-NG}}} & cassandra-0       & 10.50.103.245          & Santa Catarina        \\
\multicolumn{1}{l}{}                                   & cassandra-1       & 10.50.79.144           & Rio Grande do Sul        \\
\multicolumn{1}{l}{}                                   & cassandra-2       & 10.50.117.161          & Paraíba        \\
\multicolumn{1}{l}{}                                   & loadgen           & 10.50.83.25            & Rio Grande do Norte        \\ \hline
\multirow{4}{*}{\textbf{Fabric}}                       & cassandra-0       & 192.168.135.18          & Dallas          \\
                                                       & cassandra-1       & 192.168.104.20           & Salt Lake City           \\
                                                       & cassandra-2       & 192.168.3.78           & Lexington          \\
                                                       & loadgen           &192.168.135.13            & Dallas          \\ \hline
\end{tabular}
\end{table}

\subsection{Partial Factorial Model}\label{subsec:partial_factorial_model}

The factorial method comprises $K$ factors with $n_{i}$ levels for each $i$ factor. We used the CPU and \ ac{RAM} factors allocated to the container of each node of the Cassandra ring; for each factor, the levels were 1vCPU, 4vCPU, 2 Gb \ac{RAM}, or 8 GB RAM. Four experiments ($2^{2}$) were run on the combinations (CPU and RAM) to obtain the values $y_{1}$, $ y_{2}$, $ y_{3}$ and $y_{4}$, which are the averages of the write and read operations for each testbed. We performed an analysis using the regression model generated by the experimental combinations as follows: $y = q_{0} + q_{A}X_{A} + q_{B}X_{B} + q_{AB}X_{AB}$.

By replacing the four observations from the experiment with the model, we obtain $q_{0} = \frac{1}{4}\times(y_{1} + y_{2} + y_{3} + y{4})$ which is the average of the latencies of the operations (W and R), $q_{A} = \frac{1}{4}\times(-y_{1} + y_{2} - y_{3} + y{4})$ which is the influence of the Factor $\mathcal{A}$ (CPU) on the response variable (latency). While $q_{B} = \frac{1}{4}\times(-y_{1} - y_{2} + y_{3} + y{4})$ is the influence of the Factor $\mathcal{B}$ (RAM) on the response variable (latency), and $q_{AB} = \frac{1}{4}\times(y_{1} - y_{2} - y_{3} + y{4})$ is the influence of the $\mathcal{AB}$ Factors simultaneously on the response variable.

From the values $q_{0}$, $q_{A}$, $q_{B}$ and $q_{AB}$ we determine the sum of squares that gives the total variation of the response variables and variations in the influences of the Factors $\mathcal{A}$, $\mathcal{B}$ and $\mathcal{AB}$ simultaneously. With this, we calculated the total variance by following $SS_{T} = 2^{2}q_{A}^{2} + 2^{2}q_{B}^{2} + 2^{2}q_{AB}^{2}$. Once we have the total variance, we calculate the variance of each factor by dividing by $SS_{T}$, where the factor $SS_{A} = 2^{2}q_{A}^{2}$ is the influence of Factor $\mathcal{A}$ (CPU) on the response variable latency in operation (W and R); $SS_{B} = 2^{2}q_{B}^{2}$ which is the influence of Factor $\mathcal{B}$ (RAM) on the response variable; and finally, $SS_{AB} = 2^{2}q_{AB}^{2}$ which is the interaction of Factors $\mathcal{AB}$ (CPU and RAM) on the response variable.

\section{Results and Discussions}\label{sec:results_and_discussions}

\begin{figure}[!htb]
  \centering
  \includegraphics[width=0.5\linewidth]{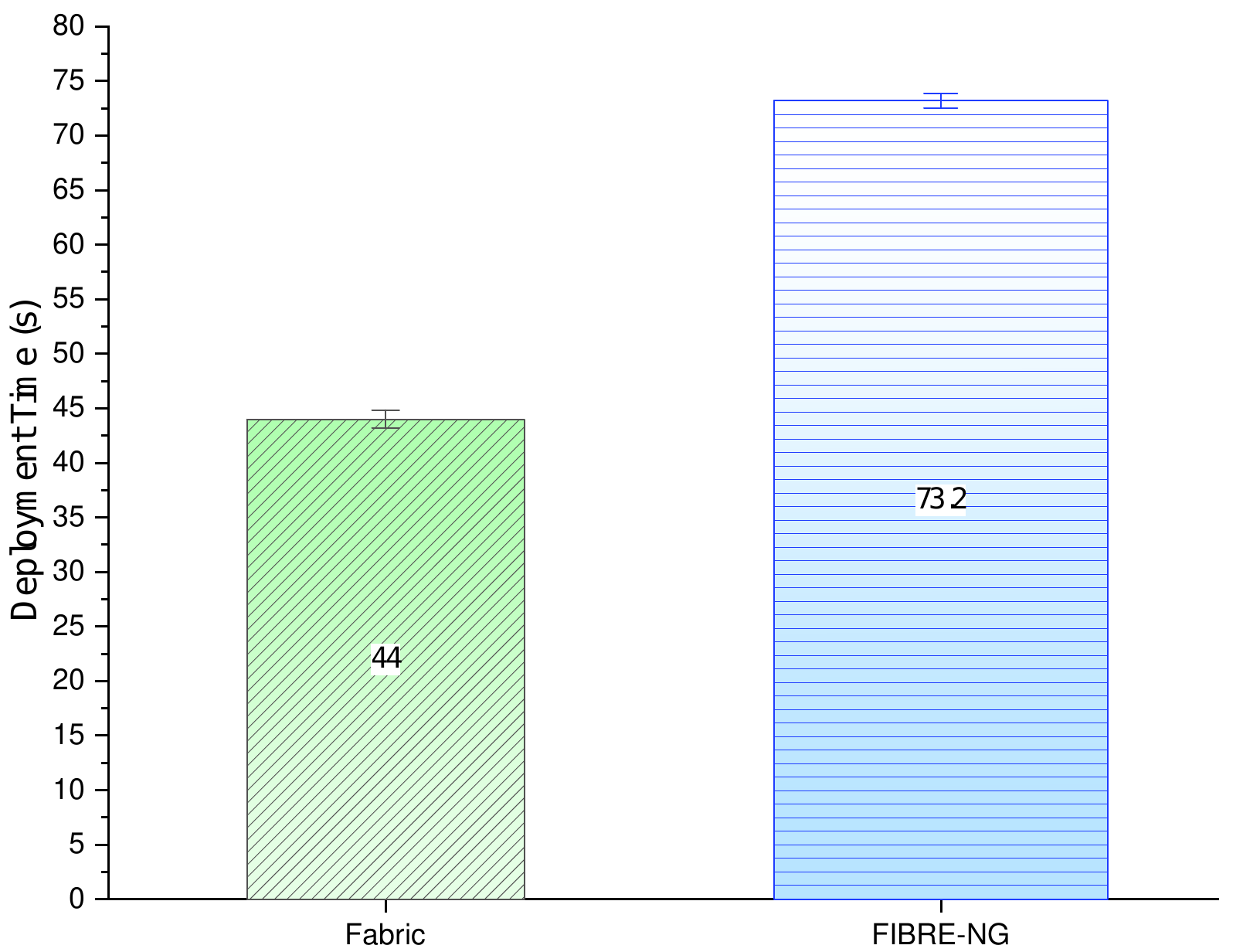}
  \caption{Deployment Time Comparison: FIBRE-NG and Fabric testbeds.}
  \label{fig:deployment_time_comparison}
\end{figure}

Initially, we measured the overhead of deploying a network slice on both testbeds, as shown in Fig. ~\ref{fig:deployment_time_comparison}. Fabric required less time to deploy the same network slice (with the template described in the manifest file). Quantitatively, FIBRE-NG required 66.36\% more time (73.2 s) to deploy the network slice than Fabric (44 s) did. This variation in deployment time may or may not be associated with heterogeneity or the amount of computational resources available to the network slice.

It is assumed that testbeds have different heterogeneous resources with variations in both hardware and software, which can lead to differences in averages. However, the aim of the experiment was to verify that identical resource allocation profiles may not lead to the same behavior for network slices on different testbeds. 

We carried out the experimental evaluation following the planned combinations (refer to Section~\ref{subsec:partial_factorial_model}) to measure the averages of the response variables for each operation (W or R). According to Table~\ref{tab:performance_evaluation_response_variables}, we observed different averages for each experimental combination in both \ac{FIBRE-NG} and Fabric. 

\begin{table}[htbp]
\centering
\caption{Network Slice performance according to resource allocation combinations on testbeds.}
\label{tab:performance_evaluation_response_variables}
\resizebox{\textwidth}{!}{%
\begin{tabular}{cccccccc}
\cline{5-8}
\multicolumn{1}{l}{} & \multicolumn{1}{l}{} & \multicolumn{1}{l}{} & \multicolumn{1}{l}{}                                                            & \multicolumn{2}{c}{\textbf{Measured on FIBRE-NG}}                                                                                              & \multicolumn{2}{c}{\textbf{Measured on Fabric}}                                                                                                \\ \hline
\textbf{Experiment}  & \textbf{CPU}         & \textbf{RAM}         & \textbf{\begin{tabular}[c]{@{}c@{}}Y: Write/Read \\ Latency  (ms)\end{tabular}} & \textbf{\begin{tabular}[c]{@{}c@{}}Write \\ Latency (ms)\end{tabular}} & \textbf{\begin{tabular}[c]{@{}c@{}}Read \\ Latency (ms)\end{tabular}} & \textbf{\begin{tabular}[c]{@{}c@{}}Write \\ Latency (ms)\end{tabular}} & \textbf{\begin{tabular}[c]{@{}c@{}}Read \\ Latency (ms)\end{tabular}} \\ \hline
\textit{\#1 }                 & 1vCPU \scriptsize{(-1)}           & 2 Gb \ac{RAM} \scriptsize{(-1)}        & $y_{1}$                                                                         & 156.9                                                                  & 100.3                                                                 & 719.2                                                                      & 616.5                                                                     \\
\textit{\#2}                  & 4vCPU \scriptsize{(1)}            & 2 Gb \ac{RAM} \scriptsize{(-1)}        & $y_{2}$                                                                         & 93.5                                                                   & 99.1                                                                  & 913.3                                                                      & 830.8                                                                     \\
\textit{\#3}                  & 1vCPU \scriptsize{(-1)}           & 8 Gb \ac{RAM} \scriptsize{(1)}         & $y_{3}$                                                                         & 186.6                                                                  & 101.3                                                                 & 404.4                                                                      & 385.8                                                                     \\
\textit{\#4}                  & 4vCPU \scriptsize{(1)}            & 8 Gb \ac{RAM} \scriptsize{(1)}         & $y_{4}$                                                                         & 93.0                                                                   & 98.2                                                                  & 265.4                                                                      & 275.8                                                                     \\ \hline
\end{tabular}%
}
\end{table}

Using the partial factorial method, we analyzed whether the allocation of CPU and \ac{RAM} to the network had similar effects on the response variable in different testbeds. In Table~\ref{tab:bench_mark_results}, we note that the CPU allocation (Factor $\mathcal{A}$) has $93.32\%$ influence on the latency of the write operation, whereas the simultaneous influence of Factors $\mathcal{B}$ and $\mathcal{AB}$ on the latency is low.

We observed a similar pattern on the same testbed: CPU allocation (Factor $\mathcal{A}$) had $83.63\%$ influence on latency. However, the simultaneous allocation of CPU and \ac{RAM} (Factor $\mathcal{AB}$) contributed $16.33\%$ to latency. This is likely due to the nature of the read operation, which includes input-output (IO) operations to access information from the input and output devices. This variation suggests that even identical resource allocation profiles, depending on the testbed, may not lead to the expected behavior of the network slice.

\begin{table}[htbp]
\centering
\caption{Resource allocation influence for network slice performance on different testbeds.}
\scriptsize
\label{tab:bench_mark_results}
\begin{tabular}{ccccc}
\cline{3-5}
                                     & \multicolumn{1}{l}{} & \multicolumn{3}{c}{\textbf{Influence}}                                              \\ \hline
\multicolumn{1}{c}{\textbf{Testbed}} & \textbf{Operation}   & \textbf{\begin{tabular}[c]{@{}c@{}}Factor $\mathcal{A}$\\ (CPU)\end{tabular}} & \textbf{\begin{tabular}[c]{@{}c@{}}Factor $\mathcal{B}$\\ (RAM)\end{tabular}} & \textbf{\begin{tabular}[c]{@{}c@{}}Factor $\mathcal{AB}$\\ (CPU and RAM)\end{tabular}} \\ \hline
\multirow{2}{*}{FIBRE-NG}            & Write                & 93.32\%                & 3.22\%                  & 3.45\%                           \\
                                     & Read                 & 83.63\%                & 0.04\%                  & 16.33\%                          \\ \hline
\multirow{2}{*}{Fabric}              & Write                & 0.29\%                    & 89.05\%                     & 10.66\%                              \\
                                     & Read                 & 1.48\%                    & 84.18\%                     & 14.34\%                              \\ \hline
\end{tabular}%
\end{table}


In the Fabric testbed, we identified the different influences on resource allocation during operations W and R. In writing, memory allocation had $89.05\%$ influence on latency, while resource interaction accounted for $10.66\%$ in response. Analysis of \texttt{cassandra-stress} indicated timeouts for the \textit{quorum} consistency level, increasing the response time, and possibly the use of \ac{RAM}. In reading, memory allocation exerted $84.18\%$ of influence, followed by $14.34\%$ of the interaction between CPU allocation and \ac{RAM} on the latency.

The influence of CPU and \ac{RAM} factors on the performance of the Cassandra application differed significantly between the two testbeds. This difference was primarily due to the high latency experienced by the slice deployed on the Fabric testbed. Although it launches network slices faster, our network slice on the Fabric testbed experienced higher latencies than the FIBER-NG testbed. This latency led to increased input–output usage and buffering in the application, which, in turn, required more \ac{RAM} allocation. In contrast, the network slice deployed on FIBRE-NG experienced lower latency, making the impact of the CPU on the application response time.

\section{Concluding Remarks}\label{sec:concluding_remarks}

This study proposes an analysis of the influence of resource allocation on network slice behavior in distributed testbeds. We followed the partial factorial model, where we combined different resource allocations for network slicing and observed the response variable (latency) for the Write and Read operations. Looking at related works, we concluded that there was an opportunity to contribute from this perspective.

After analyzing the impact of resource allocation (CPU and RAM), we concluded that although there are time differences in the deployment of the network slice in the testbeds (\ac{FIBRE-NG} and Fabric), this time does not directly influence the operation of the network slice. In addition, we found that the influence of resource allocation depends on the seasonal demand of the network slice; therefore, smart life-cycle orchestration still has opportunities.

One limitation of this study is that it focused its analysis on only two testbeds, making it difficult to generalize these results to other testbeds. In future work, we will evaluate the influence of allocating other types of resources and applying computational intelligence techniques for auto-scaling network slice resources. 

\section*{Acknowledgments}

We acknowledge the financial support of the FAPESP MCTIC/CGI Research project 2018/23097-3 - SFI2 - Slicing Future Internet Infrastructures. This study was financed in part by the Coordenação de Aperfeiçoamento de Pessoal de Nível Superior - Brasil (CAPES) - Finance Code 001. We also thank the National Council for Scientific and Technological Development (CNPq) under grant number 421944/2021-8 (call CNPq/MCTI/FNDCT 18/2021) and Centro ALGORITMI, funded by Fundação para a Ciência e Tecnologia (FCT) within the RD Units Project Scope 2020-2023 (UIDB/00319/2020) for partially support this work.

\bibliographystyle{sbc}
\bibliography{sbc-template}

\begin{thebibliography}{}

\bibitem[Arora et~al. 2024]{Arora2024}
Arora, S., Ksentini, A., and Bonnet, C. (2024).
\newblock Cloud native lightweight slice orchestration (cliso) framework.
\newblock {\em Computer Communications}, 213:1--12.

\bibitem[Baldin et~al. 2019]{baldin2019fabric}
Baldin, I., Nikolich, A., Griffioen, J., Monga, I. I.~S., Wang, K.-C., Lehman, T., and Ruth, P. (2019).
\newblock Fabric: A national-scale programmable experimental network infrastructure.
\newblock {\em IEEE Internet Computing}, 23(6):38--47.

\bibitem[Debbabi et~al. 2020]{Debbabi2020}
Debbabi, F., Jmal, R., Fourati, L.~C., and Ksentini, A. (2020).
\newblock Algorithmics and modeling aspects of network slicing in 5g and beyonds network: Survey.
\newblock {\em IEEE Access}, 8:162748--162762.

\bibitem[Donatti et~al. 2023]{Donatti2023}
Donatti, A., Correa, S.~L., Martins, J. S.~B., Abelem, A., Both, C.~B., Silva, F., Suruagy, J.~A., Pasquini, R., Moreira, R., Cardoso, K.~V., and Carvalho, T.~C. (2023).
\newblock Survey on machine learning-enabled network slicing: Covering the entire life cycle.
\newblock {\em IEEE Transactions on Network and Service Management}, pages 1--1.

\bibitem[Dong et~al. 2023]{Dong2023}
Dong, W., Li, B., Li, H., Wu, H., Gong, K., Zhang, W., and Gao, Y. (2023).
\newblock {LinkLab} 2.0: A multi-tenant programmable {IoT} testbed for experimentation with {Edge-Cloud} integration.
\newblock In {\em 20th USENIX Symposium on Networked Systems Design and Implementation (NSDI 23)}, pages 1683--1699, Boston, MA. USENIX Association.

\bibitem[Feng et~al. 2020]{Feng2020}
Feng, J., Pei, Q., Yu, F.~R., Chu, X., Du, J., and Zhu, L. (2020).
\newblock Dynamic network slicing and resource allocation in mobile edge computing systems.
\newblock {\em IEEE Transactions on Vehicular Technology}, 69(7):7863--7878.

\bibitem[Karbalaee~Motalleb et~al. 2023]{Motalleb2023}
Karbalaee~Motalleb, M., Shah-Mansouri, V., Parsaeefard, S., and Alcaraz~López, O.~L. (2023).
\newblock Resource allocation in an open ran system using network slicing.
\newblock {\em IEEE Transactions on Network and Service Management}, 20(1):471--485.

\bibitem[Moreira et~al. 2021]{Moreira2021}
Moreira, R., Rosa, P.~F., Aguiar, R. L.~A., and de~Oliveira~Silva, F. (2021).
\newblock {NASOR: A network slicing approach for multiple Autonomous Systems}.
\newblock {\em Computer Communications}, 179:131--144.

\bibitem[Morel et~al. 2023]{Morel2023}
Morel, A.~E., Calyam, P., Qu, C., Gafurov, D., Wang, C., Thareja, K., Mandal, A., Lyons, E., Zink, M., Papadimitriou, G., and Deelman, E. (2023).
\newblock Network services management using programmable data planes for visual cloud computing.
\newblock In {\em 2023 International Conference on Computing, Networking and Communications (ICNC)}, pages 130--136.

\bibitem[Padalia 2015]{padalia2015apache}
Padalia, N. (2015).
\newblock {\em Apache Cassandra Essentials}.
\newblock Packt Publishing Ltd.

\bibitem[S. et~al. 2023]{Saibharath2023}
S., S., Mishra, S., and Hota, C. (2023).
\newblock {Joint QoS and energy-efficient resource allocation and scheduling in 5G Network Slicing}.
\newblock {\em Computer Communications}, 202:110--123.

\bibitem[Salmito et~al. 2014]{salmito2014fibre}
Salmito, T., Ciuffo, L., Machado, I., Salvador, M., Stanton, M., Rodriguez, N., Abelem, A., Bergesio, L., Sallent, S., and Baron, L. (2014).
\newblock Fibre-an international testbed for future internet experimentation.
\newblock In {\em Simp{\'o}sio Brasileiro de Redes de Computadores e Sistemas Distribu{\'\i}dos-SBRC 2014}, pages p--969.

\end{thebibliography}

\end{document}